\documentclass[aps,nofootinbib,twocolumn,prl,showpacs,class-pre]{revtex4}
\usepackage{setspace} \usepackage{amsmath,amssymb,amsfonts,amsthm}
\usepackage{graphicx}

\newcommand{\be}{\begin{equation}}
\newcommand{\ee}{\end{equation}}
\newcommand{\beq}{\begin{eqnarray}}
\newcommand{\eeq}{\end{eqnarray}}

\newcommand{\Trace}{{\rm Tr}}

\def\lsim{\hbox{ \raise.35ex\rlap{$<$}\lower.6ex\hbox{$\sim$}\ }}
\def\gsim{\hbox{ \raise.35ex\rlap{$>$}\lower.6ex\hbox{$\sim$}\ }}

\begin{document}
\title{Natural inflation mechanism in asymptotic noncommutative geometry}
\author{William Nelson\footnote{william.nelson@kcl.ac.uk},
Mairi Sakellariadou\footnote{mairi.sakellariadou@kcl.ac.uk}}
\affiliation{Department of
  Physics, King's College, University of London, Strand WC2R 2LS,
  London, U.K.}

\begin{abstract}
The possibility of having an inflationary epoch within a noncommutative
geometry approach to unifying gravity and the standard model is 
demonstrated. This inflationary phase occurs without the need to
introduce {\it ad hoc} additional fields or potentials, rather it
is a consequence of a nonminimal coupling between the geometry and
the Higgs field.
\end{abstract}

\pacs{11.10.Nx, 04.50.+h, 12.10.-g, 11.15.-q, 12.10.Dm}

\maketitle

\section{Introduction}
It has long been known that despite its enormous success in explaining
the expansion of the Universe, the origin of the cosmic microwave
background radiation and the synthesis of light elements, the standard
(hot big bang) cosmological model is plagued with a number of severe
problems. More precisely, the hot big bang model is unable to address
the big bang singularity, and it cannot explain the flatness of space,
or the large-scale homogeneity and isotropy of space over causally
disconnected regions. Thus, it has to admit particular initial
conditions. In addition, it cannot explain the origin of initial
inhomogeneities giving rise to the observed structure formation,
neither can it account for the absence of dangerous relics, which
would have been formed in the early universe according to the high
energy particle physics models valid at those energy scales. Finally,
the standard cosmological model is plagued by the vacuum energy (or
cosmological constant) problem.

To address some of these shortcomings, it has been postulated that a
period of accelerated expansion, called cosmological
inflation~\cite{infl}, has proceeded the era of validity of the hot
big bang model. Inflation has not only addressed successfully the
problem of requiring particular initial conditions, but it has also
succeeded in predicting the (almost) scale invariant spectrum of
density perturbations that are measured in the cosmic microwave
background temperature anisotropies. Although some questions have yet
to be fully answered, such as the specifics of
reheating~\cite{Graham:2008vu} and the likelihood of the onset of
inflation~\cite{Calzetta:1992gv}, the remarkable agreement with
observation makes inflation an extremely attractive approach to
understanding the early universe.

Unfortunately, it has proved difficult to naturally embed inflation
within an underlying fundamental theory.  Inflation most naturally
occurs when the dynamics of the universe are dominated by the
evolution of a scalar field, the inflaton, slowly rolling in its
potential; the form of the potential defines the type of the
inflationary model. There is only one scalar field within the standard
model of particle physics, the Higgs field, and it is naturally hoped
that this could play the r\^ole of the inflaton. However, it has been
shown~\cite{Bezrukov:2008ej} that in order for the Higgs field to
produce the correct amplitude of density perturbations, its mass would
have to be some $11$ orders of magnitude higher than the one required
by particle physics.  This conclusion was however reached using
general relativistic cosmology and here we re-examine the calculation
in the context of noncommutative geometry~\cite{Connes:1994yd}.

\section{Noncommutative geometry}
\subsection{Motivation}

Despite the long efforts, a unified theory of all interactions,
including gravity, remains still lacking. The reason for this
difficulty may have its origin in the different properties and
underlying symmetries of the Einstein-Hilbert Lagrangian ${\cal
  L}_{\rm EH}$, and the Standard Model (SM) Lagrangian ${\cal L}_{\rm
  SM}$.  Certainly, for physical processes much below the Planck scale
($\approx 10^{18}$ GeV), gravity can be safely considered as a
classical theory. However, as energies approach the Planck scale, the
quantum nature of space-time becomes apparent, and the
Einstein-Hilbert action becomes an approximation. Moreover, at Planck
scale, one expects all forces of nature (including gravity) to become
unified.  The structure of space-time at Planckian energies is one of
the fundamental unanswered questions in physics today.  At such
scales, geometry can no longer be described in terms of the Riemannian
geometry and General Relativity; one should search for a paradigm of
geometry within the quantum framework. Such an attempt has been
realised within the concept of NonCommutative Geometry (NCG).

To be more precise, considering the SM minimally coupled to gravity,
the physical laws at sufficiently low energies can be described by the
sum ${\cal L}= {\cal L}_{\rm EH}+{\cal L}_{\rm SM}$.  The symmetry
group of ${\cal L}_{\rm EH}$ is, by the equivalence principle, the
diffeomorphism group, ${\rm Diff}({\cal M})$, of the space-time
manifold ${\cal M}$. However, the symmetry of the gauge theory in
${\cal L}_{\rm SM}$, is the group of local gauge transformations ${\rm
  G}_{\rm SM}=C^{\infty}({\cal M},{\rm U}(1)\times {\rm SU}(2)\times
{\rm SU}(3))$.  Thus, considering the Lagrangian ${\cal L}$, the full
symmetry group ${\rm G}$ will be a semidirect product ${\rm G}({\cal
  M}) = {\rm G}_{\rm SM}({\cal M}) \rtimes {\rm Diff}({\cal M})$. To
argue that the whole theory is pure gravity on a space ${\cal M}$, one
should find such a space for which ${\rm G}={\rm Diff}({\cal
  M})$. However, it is not possiblt to find such a space among
ordinary manifolds, instead one needs to consider noncommutative
spaces.  The noncommutative space is a product ${\cal M}\times {\cal
  F}$ of an ordinary space-time manifold ${\cal M}$, by a finite ({\sl
  i.e.}, the algebra of coordinates on ${\cal M}$ is finite
dimensional) noncommutative space ${\cal F}$.

To extract physical applications of NCG we will use its main idea,
namely that all information about a physical system is contained
within the algebra of functions, represented as operators in a Hilbert
space, while the action and metric properties are encoded in a
generalised Dirac operator.  We will then look for a geometry (in the
noncommutative sense, {\sl i.e.}, by specifying an algebra ${\cal A}$,
a Hilbert space ${\cal H}$ and a generalised Dirac operator $D$), such
that the associated action functional produces the SM of
electroweak and strong interactions with all its refinements
prescribed by experimental data~\footnote{The self-adjoint operator in
  a Hilbert space ${\cal H}$, is the quantum analogue of the classical
  real variable.  More precisely, complex and real variables,
  differentials and integrals have they corrsponding analogues in the
  {\sl quantised} calculus dictated by the noncommutative differential
  geometry.}.  

There is a very simple noncommutative algebra ${\cal A}$, whose group
of inner automorphisms~\footnote{Corresponding in physics to {\sl
    internal symmetries}.} corresponds to the group of gauge
transformations ${\rm G}_{\rm SM}({\cal M})$, and it has a quotient
that corresponds exactly to diffeomorphisms~\cite{Connes:2006ms}.  The
noncommutative algebra ${\cal A}$ is a direct sum $\mathbb{C}\oplus
\mathbb{H}\oplus M_3(\mathbb{C})$, with $\mathbb{C}, \mathbb{H},
M_3(\mathbb{C})$ denoting the algebra of complex numbers, quaternions,
and $3\times 3$ complex matrices, respectively. The algebra ${\cal A}$
corresponds to a finite space where the SM fermions and the Yukawa
parameters determine the spectral geometry. The Hilbert space ${\cal
  H}$ is finite dimensional and admits the set of elementary fermions
as a basis. The fermionic fields acquire mass through the spontaneous
symmetry breaking produced by the Higgs field.  The Standard Model of
elementary particle physics provides an extraordinary example of a
{\sl spectral triple}~\footnote{The spectral triple $({\cal A, H}, D)$
  encodes the geometry, given as a Hilbert space representation of the
  pair $({\cal A}, D)$.} in the noncommutative
setting~\cite{Chamseddine:2006ep}. The exciting outcome of this theory
is that the Higgs appears naturally as the ``vector'' boson of the
internal noncommutative degrees of freedom.

In the past, the connection between strings and NCG has been investigated,
while more recently connections between NCG and Loop Quantum
Gravity are emerging. Given the plethora of very precise high energy
physics data from astroparticle and cosmology, possibly also combined
with the Large Hadron Collider, the geometry of space and the laws of
physics at the Planck energy scale will not remain a mystery for much
longer.

\subsection{Elements of NCG}

Consider the extension of our smooth four-dimensional manifold ${\cal
  M}$, by taking the product of it with a discrete noncommuting
manifold ${\cal F}$ of $KO$-homology dimension ({\sl i.e.}, the
dimension modulo 8) equal ~\footnote{The Standard Model with neutrino
  mixing favors the shift of dimension from the (familiar) 4 to
  $10=4+6=2 \ {\rm modulo} 8$~\cite{Chamseddine:2008zj}.} to $6$. This
internal space has dimension $6$ to allow fermions to be
simultaneously Weyl and chiral (as within string theory), whilst it is
discrete to avoid the infinite tower of massive particles that are
produced in string theory.  The noncommutative nature of ${\cal F}$ is
given by a spectral triple $\left( {\cal A},{\cal H}, D\right)$, where
${\cal A}$ is an involution of operators on the Hilbert space ${\cal
  H}$, which is essentially the algebra of coordinates, and $D$ is a
self-adjoint unbounded operator~\footnote{The operator $D$ has a
  direct physical meaning; it is given by the Yukawa coupling matrix
  which encodes the masses of the 9 elementary fermions as well as the
  4 mixing parameters of the Standard Model.} in ${\cal H}$, such that
all commutators $\left[ D,a \right]$ are bounded for $a\in{\cal A}$,
and $(D-\lambda)^{-1}$ is compact for any $\lambda \notin \mathbb{R}$.
The operator $D$ corersponds to the inverse line element of Riemannian
geometry, whilst the commutator $[D,a],a\in {\cal A}$ will play the
r\^ole of the differential quotient ${\rm d}a/{\rm d}s$, with ${\rm
  d}s$ the unit of length.

By assuming that the algebra constructed in ${\cal M}\times {\cal F}$
is {\it symplectic-unitary}, the algebra ${\cal A}$ is restricted to
be of the form $\mathcal{A}=M_{a}(\mathbb{H})\oplus
M_{k}(\mathbb{C})$, where $k=2a$. The choice $k=4$ is the first value
that produces the correct number of fermions in each
generation~\cite{Chamseddine:2007ia} (note however that the number of
generations is an assumption in the theory). Finally, the Dirac
operator $D$ connects ${\cal M}$ and ${\cal F}$ via the action
functional, called {\sl spectral action}, of the form ${\rm Tr}\left(
f\left(D/\Lambda\right)\right)$, where $f$ is a test function (a
smooth even function with fast decay at infinity) and $\Lambda$ is the
cut-off energy scale, introduced so that $D/\Lambda$ becomes
dimensionless~\footnote{It accounts only for the bosonic part of the
  model. The coupling with fermions is obtained by including an
  additional term, namely $ {\rm Tr}(f(D/\Lambda))+(1/2)\langle
  J\psi,D\psi\rangle$, with $J$ the real structure on the spectral
  triple, and $\psi$ an element in the space ${\cal H}$, viewed as a
  classical fermion~\cite{Chamseddine:2006ep}.} The expression ${\rm
  Tr}\left( f\left(D/\Lambda\right)\right)$ is taken as a natural
spectral formulation of gravity, while it can be also used for spaces
which are not Riemannian, and in particular for our choice of ${\cal
  M}\times {\cal F}$. Moreover, the spectral action has theee main
advantages. Firstly, when $f$ is a cut-off function (so, $f\geq 0$) ,
the spectral action is just counting the number of eigenstates of $D$
in the interval $[-\Lambda,\Lambda]$, and secondly ${\rm Tr}\left(
f\left(D/\Lambda\right)\right)\geq 0$, namely it has the correct sign
for a Euclidean action. Thirdly, the functional ${\rm Tr}\left(
f\left(D/\Lambda\right)\right)$ is invariant under the unitary group
of the Hilbert space ${\cal H}$.

Asymptotically, it can be shown~\cite{Chamseddine:2006ep} that this
approach leads to an effective four dimensional action that includes
all the standard model particles, with the correct couplings,
including the right-handed neutrinos as well as the {\sl see-saw}
mechanism. The gravitational and Higgs part of this action
read~\cite{Chamseddine:2006ep} 
\beq
\label{eq:action1} 
{\cal S}_{\rm grav} = \int \left( \frac{1}{2\kappa_0^2} R + \alpha_0
C_{\mu\nu\rho\sigma}C^{\mu\nu\rho\sigma} + \tau_0 R^\star
R^\star\right.  \nonumber\\ \left. +\gamma_0 -\xi_0 R|{\bf H}|^2
+\frac{1}{2} |D_\mu {\bf H}|^2 + V\left( |{\bf H}|\right) \right)
\sqrt{g} {\rm d}^4 x~, 
\eeq 
where ${\bf H}$ is the Higgs field, normalised to have a canonical
kinetic term, the potential $V\left(|{\bf H}|\right)= \lambda_0|{\bf
  H}|^4 - \mu_0^2 |{\bf H}|^2$, is the standard Higgs potential and
the $\kappa_0^2, \alpha_0, \tau_0, \lambda_0, \mu_0$ are specified in
terms of the cut-off energy scale $\Lambda$, the couplings $a,b,c,d,e$,
given by~\cite{Chamseddine:2006ep} 
\begin{eqnarray}
\label{couplings}
  a &=& \,\Trace({Y}_{ \uparrow 1}^*{Y}_{ \uparrow 1}+
  {Y}_{\downarrow 1}^* {Y}_{\downarrow 1} +3({Y}_{\uparrow 3}^*{Y}_{\uparrow
    3}+{Y}_{\downarrow 3}^*{Y}_{\downarrow 3}))~,
  \nonumber \\
  b &=& \,\Trace(( {Y}_{\uparrow 1}^*{Y}_{ \uparrow
    1})^2+({Y}_{\downarrow 1}^*{Y}_{\downarrow 1})^2+3({Y}_{\uparrow
    3}^*{Y}_{\uparrow 3})^2\nonumber\\
&&~~~~~~~~~~~~~~~~~~~~~~~~~~~~~~~~~ +3({Y}_{\downarrow
    3}^*{Y}_{\downarrow 3})^2)~,
\nonumber \\
 c &=&  \Trace({Y}_R^*{Y}_R)~,
\nonumber \\
d &=&  \Trace(({Y}_R^*{Y}_R)^2)~,
 \nonumber \\
e &=& \Trace({Y}_R^*{Y}_R {Y}_{\uparrow 1}^*{Y}_{\uparrow 1})~,
\end{eqnarray}
and the coefficients $f_k=\int_0^\infty f(v)v^{k-1} {\rm d}v$ for
$k>0$ which is related to the coupling constants at unification and
allows the action of the quaternions ${\mathbb H}$ to be expressed in
terms of Pauli matrices as $q=f_0 + \sum i f_\alpha \sigma^\alpha$.
Note that the $Y$'s are used to classify the action of the Dirac
operator and give the fermion and lepton masses, as well as lepton
mixing, in the asymptotic version of the spectral action. The value of
the coupling $\xi_0$ is set $\xi_0=1/12$. 
The couplings $a, \dots, e$
are determined by the (unimodular) inner fluctuations of the metric.

In Ref.~\cite{Nelson:2008uy} we have shown that the equations of
motion of the gravitational part of Eq.~(\ref{eq:action1}), in a
homogeneous and isotropic background are exactly those of standard
general relativity. Thus, background cosmology remains unchanged
within this noncommutative approach to the standard model. We
emphasise that this is the effect of the purely geometrical terms; the
term $R^\star R^\star$ is topological and hence plays no r\^ole in
dynamics, while the term $C_{\mu\nu\rho\sigma}C^{\mu\nu\rho\sigma}$
vanishes for homogeneous and isotropic metrics. Thus, we are left only
with the standard Einstein-Hilbert term. It is important to remember
however that inhomogeneous perturbations to this background will
evolve differently from the equivalent classical system.

Equation (\ref{eq:action1}) implies that, in addition to the
cosmological constant term $\gamma_0$, which we neglect here, the
geometry is nonminimally coupled to the Higgs field. In what follows,
we investigate the consequences of this nonminimal coupling, with
respect to the possibility of having naturally an inflationary
scenario driven by the Higgs field.  Remarkably, such modifications to
the Einstein-Hilbert gravity have already been recently considered in
the literature~\cite{Bezrukov:2008ej,Bezrukov:2007ep}. In those
studies, the nonminimal coupling was postulated, and then shown that
the scale that sets the amplitude of perturbations during Higgs driven
inflation is $\lambda_0/\xi_0^2$, rather than simply $\lambda_0$ as is
the case without this additional nonminimal coupling. Indeed, this
reduction in the amplitude of induced perturbations allows this Higgs
field to satisfy the requirements of the standard model, as well as
inflation simultaneously.

To be more specific, in Ref.~\cite{Bezrukov:2008ej} a conformal
transformation of the metric was used, such that
\be
 \left( \frac{1}{2\kappa_0^2} -\xi_0 |{\bf H}|^2 \right) R \ \rightarrow \
-\frac{1}{2\kappa_0^2} \hat{R}~.
\ee
This leads to a noncanonical kinetic term for $|{\bf H}|$ which is
removed via a re-definition of the field $|{\bf H}| \rightarrow |\chi|$
to give the {\it Einstein frame} action
\be
 {\cal S}_{\rm E} = \int \left( -\frac{1}{2\kappa_0^2} \hat{R} +
\frac{1}{2} |D_\mu \chi||D^\mu \chi| - U\left(\chi\right) \right) \sqrt{g}
{\rm d}^4 x~,
\ee
where in the limit $|{\bf H}| \gg (\kappa_0\sqrt{2\xi_0})^{-1}$,
the potential $U\left(\chi\right)$, is given by
\be
 U\left(\chi\right) \approx \frac{ \lambda_0}{4\kappa_0^4 \xi_0^2}
\left[ 1- \exp \left( -\frac{2\chi_0}{\sqrt{6}\kappa_0
} \right)\right]^{2}~.
\ee
It is the flatness of this potential that allows slow-roll inflation
to occur.  The above employed conformal transformation allows the
system to be analysed in a standard manner. Note however that the
effects of such a nonminimal coupling between the geometry and the
Higgs field have been also investigated directly in the {\it Jordan
  frame}, i.e., without doing the conformal
transformation~\cite{Tsujikawa:2004my}.
 
Normalising the cosmic microwave background perturbations to the
WMAP5 data~\cite{Komatsu:2008hk}, implies the requirement
\be 
\xi_0 \approx 44700 \sqrt{\lambda_0}~, 
\ee 
which ensures that the Higgs field can produce inflation. Moreover,
the spectral index $n_s \approx 0.97$ and the tensor-to-scalar ratio
$r \approx 0.003$, are well within the WMAP5 limits. This conclusion is
maintained under tree level~\cite{Bezrukov:2007ep} and
one-loop~\cite{Bezrukov:2008ej} running of the couplings, provided
the Higgs mass is in the, experimentally viable, range 
\be 
136.7 \ {\rm GeV} < m_{\rm H} <  184.5 \ {\rm GeV} \  {\rm
  for}\ \ m_{\rm top} = 171.2 \ {\rm GeV}~.
\ee
Note that two-loop calculations may lead to significant effects on the
running of the Higgs potential~\cite{Malcolm,DeSimone:2008ei}.

In the context of the noncommutative approach to the SM
however, the couplings $\xi_0$ and $\lambda_0$ are not arbitrary.
Namely, since the action, Eq.~(\ref{eq:action1}), comes from an 
underlying theory, we have some control on the values of the couplings
$\xi_0$ and $\lambda_0$. More precisely,
\be
 \xi_0 = \frac{1}{12}\ \ 
\mbox{and}~~ \lambda_0 = \frac{\pi^2}{2f_0}\frac{b}{a^2}~.
\ee
Hence, within the noncommutative approach to the SM, for
inflation to be naturally viable {\it without the need to introduce
  additional nonstandard model fields}, we need
\be
\label{req-infl}
\frac{b}{f_0a^2} \approx 7.04 \times 10^{-13}~,  
\ee 
where $a, b$ are defined in Eq.~(\ref{couplings}) and $f_0=f(0)$, with
$f_k$ defined as previously.

A detailed analysis of the running of these values with the cut-off
scale would determine the energy scale at which inflation occurred.
More precisely, one should compare the requirements,
Eq.~(\ref{req-infl}), so that the Higgs field can play the r\^ole of
the inflaton, with those stemming from the particle physics
phenomenology of the SM. Unfortunately, the restrictions of the
running of the couplings, found in the
literature~\cite{Chamseddine:2006ep}, have neglected the nonminimal
coupling of the Higgs to the geometry, which as we have seen is indeed
crucial for the inflation to be successful. Nevertheless, a back of
the envelope calculation shows that since $b/a^2\geq
1/4$~\cite{Chamseddine:2006ep}, a relation which is valid even for a
large tau neutrino Yukawa coupling, Eq.~(\ref{req-infl}) implies a
severe constraint on $f_0$. Alternatively, since $(g_3^2
f_0)/(2\pi^2)=(1/4)$, one obtains equivalently a constraint on the
gauge coupling $g_3$.  Since $g_3^2=g_2^2=(5/3) g_1^2$ and at the
unification scale $\Lambda\sim 1.1\times 10^{17} {\rm GeV}$, the three
coupling constants are $\alpha_i(\Lambda)=g_i^2/(4\pi)$, we obtain \be
f_0={\pi\over 8\alpha_2(\Lambda)}\sim 18.45~, \ee which does not
satisfy the requirement, Eq.~(\ref{req-infl}), so that the Higgs field
can play the r\^ole of the inflaton.  More precisely, the constraint
on $f_0$ so that inflation can be naturally incorporated here, reads
$f_0\sim 3.55\times 10^{11}$~.

Higgs inflation in the context of {\sl conventional} cosmological
models (as {\sl e.g.}, \cite{Bezrukov:2008ej,Bezrukov:2007ep}) has
been criticised~\cite{Burgess:2009ea}, arguing that quantum
corrections to the semi-classical approximation may no longer be
neglected for such {\sl exotic} inflationary models. However, this
criticism is not applicable to the noncommuative approach employed
here. More precisely, in {\sl conventional} Higgs inflation there is a
strong coupling, namely $\xi_0\sim 10^4$ between the Higgs field and the
Ricci carvature scalar. Thus, the effective theory ceases to be valid
beyond a cut-off scale $m_{\rm Pl}/\xi_0$, while one should know the
Higgs potential profile for the field values relevant for inflation,
namely $m_{\rm Pl}/\sqrt\xi_0$, values which is much bigger than the
cut-off. Clearly, this arguement does not apply to the noncommutative
Higgs drive inflation, since there $\xi_0=1/12$.

\section{Conclusions}

Considering the product of ordinary Euclidean space-time ({\sl i.e.},
space-time but with imaginary time) by a finite space (with the
properties discussed above), a geometric interpretation of the
experimentally confirmed effective low energy model of particle
physics was given in Ref.~\cite{Chamseddine:2006ep}. 

Investigating cosmological consequnecs of this proposal, we have
concluded that the Higgs field can play the r\^ole of the inflaton
field within the noncommutative approach to the standard model,
provided inflation will take place at a scale higher than the strong
weak unification scheme, $10^{17} {\rm GeV}$. In order to find the
precise value of this scale, a detailed analysis of the running of the
couplings above unification would be required. However, let us
emphasise that the aim of this paper is simply to note that within the
noncommutative geometry approach to unifying gravity and the Standard
Model, it is possible to have an epoch of inflation sourced by the
dynamics of the Higgs field.
            
In addition, this type of {\sl noncommutative inflation} could have
specific consequences that would discriminate it from alternative
models. In particular, since the theory contains all of the Standard
Model fields, along with their couplings to the Higgs field, which in
this scenario plays the r\^ole of the inflaton, a quantitative
investigation of reheating should be possible. More significantly, the
cosmological evolution equations for inhomogeneous perturbations
differs from those of the standard
Friedmann-Lema\^{i}tre-Robertson-Walker
cosmology~\cite{Nelson:2008uy}. This raises the possibility that
signatures of this {\sl noncommutative inflation} could be contained
within the cosmic microwave background power spectrum.

\vskip.2truecm 
\acknowledgements 
It is a pleasure to thank Ali Chamseddine and Malcolm Fairbairn for
discussions.  The work of M.S. is partially supported by the European
Union through the Marie Curie Research and Training Network {\sl
  UniverseNet} (MRTN-CT-2006-035863).

\end{document}